AGILE-CYG-X-3_Nat_012

August 7, 2009

SUBMITTED TO NATURE

# Discovery of extreme particle acceleration in the microquasar Cygnus X-3

M. Tavani[1,2,3,4], A. Bulgarelli[5], G. Piano[1,4], S. Sabatini[2,4], E. Striani[2,4], Y. Evangelista[1], A. Trois[1], G. Pooley[6], S. Trushkin[7], N.A. Nizhelskij[7], M. McCollough[8], K.I.I. Koljonen[9], G. Pucella[10], A. Giuliani[11], A.W. Chen[3,11], E. Costa[1], V. Vittorini[1,3], M. Trifoglio[5], F. Gianotti[5], A. Argan[1], G. Barbiellini[3,12,13], P. Caraveo[11], P.W. Cattaneo[14], V. Cocco[2], T. Contessi[11], F. D'Ammando[1,2], E. Del Monte[1], G. De Paris[1], G. Di Cocco[5], G. Di Persio[1], I. Donnarumma[1], M. Feroci[1], A. Ferrari[3,15], F. Fuschino[5], M. Galli[16], C. Labanti[5], I. Lapshov[17], F. Lazzarotto[1], P. Lipari[18,19], F. Longo[12,13], E. Mattaini[11], M. Marisaldi[5], M. Mastropietro[20], A. Mauri[5], S. Mereghetti[11], E. Morelli[5], A. Morselli[4], L. Pacciani[1], A. Pellizzoni[21], F. Perotti[11], P. Picozza[2,4], M. Pilia[21,22], M. Prest[22], M. Rapisarda[10], A. Rappoldi[14], E. Rossi[5], A. Rubini[1], E. Scalise[1], P. Soffitta[1], E. Vallazza[13], S. Vercellone[23], A. Zambra[3,11], D. Zanello[18,19], C. Pittori[24], F. Verrecchia[24], P. Giommi[24], S. Colafrancesco[24], P. Santolamazza[24], A. Antonelli[25], L. Salotti[26]

1. *INAF-IASF Roma, via del Fosso del Cavaliere 100, I-00133 Roma, Italy*
2. *Dipartimento di Fisica, Università Tor Vergata, via della Ricerca Scientifica 1,I-00133 Roma,  Italy*
3. *Consorzio Interuniversitario Fisica Spaziale (CIFS), villa Gualino - v.le Settimio Severo 63, I-10133 Torino, Italy*
4. *INFN Roma Tor Vergata, via della Ricerca Scientifica 1, I-00133 Roma, Italy*
5. *INAF-IASF Bologna, via Gobetti 101, I-40129 Bologna, Italy*
6. *Astrophysics Group, Cavendish Lab., 19 J.J.Thomson Avenue, Cambridge CB3 0HE, UK*
7. *Special Astrophysical Observatory RAS,Karachaevo-Cherkassian res., Nizhnij Arkhyz, 36916, Russia*
8. *Smithsonian Center for Astrophysics, 60 Garden Street, Cambridge, MA 02138, USA*
9. *TKK/Metsähovi Radio Observatory, Metsähovintie 114, Kylmälä 02540, Finland*
10. *ENEA Frascati, via Enrico Fermi 45, I-00044 Frascati (RM), Italy*
11. *INAF-IASF Milano, via E. Bassini 15, I-20133 Milano, Italy*
12. *Dip. Fisica, Università di Trieste, via A. Valerio 2, I-34127 Trieste, Italy*
13. *INFN Trieste, Padriciano 99, I-34012 Trieste, Italy*
14. *INFN Pavia, via Bassi 6, I-27100 Pavia, Italy*
15. *Dipartimento di Fisica, Università di Torino, via P. Giuria1, I-10125 Torino, Italy*
16. *ENEA Bologna, via don Fiammelli 2, I-40128 Bologna, Italy*
17. *IKI, Profsoyuznaya Street 84, Moscow 117997, Russia*
18. *INFN Roma 1, p.le Aldo Moro 2, I-00185 Roma, Italy*
19. *Dip. Fisica, Università La Sapienza, p.le Aldo Moro 2, I-00185 Roma, Italy*
20. *CNR, IMIP, I-00016 Montelibretti (Rome), Italy*
21. *INAF Osservatorio Astron. Cagliari, Poggio dei Pini, I-09012 Capoterra, Italy*
22. *Dip. Fisica, Univ. Insubria, via Valleggio 11, I-22100, Como, Italy*
23. *INAF IASF Palermo, via La Malfa 153, I-90146 Palermo, Italy*
24. *ASI Science Data Center, ESRIN, I-00044 Frascati, Italy*
25. *Osservatorio Astronomico di Roma, via di Frascati 33, I-00040 Monte Porzio Catone, Italy*
26. *Agenzia Spaziale Italiana, viale Liegi 26, I-00198Roma, Italy*

**The study of relativistic particle acceleration is a major topic of high-energy astrophysics. It is well known that massive black holes in active galaxies can release a substantial fraction of their accretion power into energetic particles, producing gamma-rays and relativistic jets. Galactic microquasars[1] (hosting a compact star of 1-10 solar masses which accretes matter from a binary companion) also produce relativistic jets. However, no direct evidence of particle acceleration above GeV energies has ever been obtained in microquasar ejections, leaving open the issue of the occurrence and timing of extreme matter energization during jet formation. Here we report the detection of transient gamma-ray emission above 100 MeV from the microquasar Cygnus X-3, an exceptional X-ray binary[2,3,4,5,6] which sporadically produces powerful radio jets[7,8,9]. Four gamma-ray flares (each lasting 1-2 days) were detected by the AGILE satellite[10,11] simultaneously with special spectral states of Cygnus X-3 during the period mid-2007/mid-2009. Our observations show that very efficient particle acceleration and gamma-ray propagation out of the inner disk of a microquasar usually occur a few days *before* major relativistic jet ejections. Flaring particle energies can be thousands of times larger than previously detected maximum values (with Lorentz factors of $10^5$ and $10^2$ for electrons and protons, respectively). We show that the *transitional nature* of gamma-ray flares and particle acceleration above GeV energies in Cygnus X-3 is clearly linked to special radio/X-ray states preceding strong radio flares. Thus gamma-rays provide unique insight into the nature of physical processes in microquasars.**

Cygnus X-3 (Cyg X-3) is a powerful X-ray binary of period[3] $P_{orb}$=4.8 hr and typical luminosity near the maximum accretion power of a solar mass compact star[12], $L_X \approx 10^{38}$ erg $s^{-1}$, for a 10 kpc distance[5]. The compact star powering the system is either a neutron star in an unusual state of accretion, or a black hole of 10-20 solar masses orbiting around a Wolf-Rayet companion[13]. Thus far, emission up to ~300 keV has been detected[14,15] from Cyg X-3, that usually shows a complex X-ray spectrum with two main states ("soft" and "hard") .

We report here the results of extensive observations of Cyg X-3 at gamma-ray energies by the AGILE satellite during mid-2007/mid-2009. The AGILE gamma-ray instrument[10,11] is very compact and capable of monitoring cosmic sources simultaneously in the gamma-ray (100 MeV – 10 GeV) and in the hard X-ray energy band (18-60 keV) with good sensitivity and optimal angular resolution. AGILE is operating in a fixed-pointing mode, implying that it can accumulate data on a source within its large field of view (2.5 sr for the gamma-ray imager) fourteen times a day, reaching a 1-day exposure above 100 MeV of $10^7$ $cm^2$ sec. This capability is ideal for detecting microquasar short timescale variability (down to a few hours).

AGILE accumulated on Cyg X-3 a total exposure above 100 MeV near $10^9$ $cm^2$ sec (equivalent to an effective duration of about 5 months). The gamma-ray intensity map of the Cygnus region is shown in Fig. 4 (Supplementary Information). This region is complex, hosting star formation sites, OB associations, and several prominent X-ray sources. Galactic diffuse gamma-ray radiation is taken into account by modelling[16] the interaction of cosmic rays with interstellar matter and radiation fields along the line of sight. The good AGILE angular resolution satisfactorily resolves the field surrounding Cyg X-3 at gamma-ray energies. A dominant source positioned at 0°.4 from the Cyg X-3 position is the steady gamma-ray pulsar[17] 1AGL J2032+4102/0FGL J2032.2+4122. By integrating all AGILE data, we find a weak (4.6 sigma) gamma-ray source consistent with the Cyg X-3 position and flux $F = (10 \pm 2) \cdot 10^{-8}$ photons $cm^{-2}$ $s^{-1}$ above 100 MeV (see Suppl. Information).

In addition, we also detect *transient* gamma-ray emission from a flaring source consistent with the Cyg X-3 position in four distinct episodes reported in Table 1 (shown in Fig. 5 of Suppl. Information). These flares were found by an independent multi-source maximum likelihood search for transients in all available AGILE data (off-axis angles less than $45^o$, covering an area of $5^o$x$5^o$ centered at the Cyg X-3 position). The statistical significance of all flares was individually assessed by both the maximum likelihood analysis and False Discovery Rate[18,19] method. They all passed stringent post-trial significance requirements, implying an occurrence rate for equivalent statistical fluctuations larger than 1 over several hundred 1-day map replications (see Suppl. Information). We obtain flux, significance, and positioning of these events as reported in Table 2 (Suppl. Information).
By integrating all flaring data, we find a transient gamma-ray source detected at 5.5 sigma level at the average Galactic coordinate location (l,b): (79º.6, 0º.5) ± 0º.5 (stat.) ± 0º.1 (syst.), with average flaring flux F = (190 ± 40) x $10^{-8}$ photons $cm^{-2}$ $s^{-1}$ above 100 MeV. After careful investigation, we attribute this flaring source to Cyg X-3: no other prominent source is known within the AGILE gamma-ray error box. The average spectrum between 100 MeV and 3 GeV is well described by a power-law of photon index 1.8 ± 0.2.

These flares are all associated with quite special Cyg X-3 radio and X-ray/hard X-ray states. Fig. 1 shows the Cyg X-3 daily flux lightcurve at hard X-ray energies (15-50 keV) as monitored by the *Swift*-BAT instrument during Jan. 1, 2008 – Jun. 30, 2009. Remarkably, *all* gamma-ray flares in Table 1 (marked by red arrows in Fig. 1) occur during distinct *minima* of the hard X-ray lightcurve. As well known[6,14], the Cyg X-3 hard (20 keV and above) and regular X-ray (1-10 keV) fluxes are anticorrelated. Gamma-ray flares occur then only during *soft X-ray states* or their transitions to or from quenched hard X-ray states.

A further crucial piece of information is provided by the Cyg X-3 radio states as monitored by our group with the AMI Large Array[25] and RATAN-600 radio telescopes. We find that 3 out of 4 gamma-ray flares are distinctively produced *before* major radio flares. Fig. 2 shows the radio, gamma-ray, soft X-ray, and hard X-ray energy data of the strongest radio flare of our sample (April 18, 2008 when Cyg X-3 reached the flux level of 16 Jy at 11 GHz; refs. 20,21) . Fig. 7 in the Supplementary Information show the multifrequency data of the Dec. 12, 2008 gamma-ray flare also followed by a very strong radio outburst. We find that in 2 out of 4 cases (2-3 Nov., 2008, and 21-22 June 2009, see Fig. 6 and Fig. 8) the gamma-ray activity is associated with the source entering into a “quenched radio state” (ref. 22), i.e., a rare state (2% of the time) of distinct *minimum* of the radio flux that usually anticipates a major radio flare[8,14].

Our results show clearly that the flaring gamma-ray emission occurs only at special *transitional states*, which are associated with bright soft X-ray states and/or the very low radio emission which precedes major radio flares. In order to clarify this point, we find particularly enlightening to update with our data the schematic view of the radio/X-ray transitions as done in ref. 14. Fig. 3 shows the peculiar “hysteresis curve” that Cyg X-3 follows in a radio/soft X-ray diagram representing all spectral states of the source. Very remarkably, the gamma-ray flares detected by AGILE tend to occur at the special "gully" of the hysteresis curve corresponding to very low radio states/strong soft X-ray states as preludes to major radio flares. Cyg X-3 spends only a few percent of its time in these rare states[8,14].

Our detection of transient gamma-ray emission above 100 MeV from Cyg X-3 provides the first direct evidence that extreme particle acceleration and non-thermalized emission can occur in microquasars with a repetitive pattern. Thus far, the complex interplay between the inner disk (and presumably soft X-ray) emission and coronal or jet emission in Cyg X-3 (most likely related with hard X-ray emission) has been addressed by Comptonization models that reproduce[14,15,23] the spectral states up to ~300 keV. Despite a number of important differences with other Galactic systems[23], the Cyg X-3 states resemble those of accreting black holes with a mixed population of thermal and non-thermal relativistic electrons. The Cyg X-3 X-ray spectral states are fitted by optically thick Comptonized models with the important inclusions of a reflection component[24] and a high-energy power-law additional component[14,15]. Typical temperatures of the hot coronal plasma reach ~10 keV, and the high-energy tail has been modelled to extend up to photon energies of an MeV or slightly higher[14,15].

However, Cyg X-3 is capable of accelerating particles by a very efficient mechanism leading to photon emission at energies thousands of times larger than the maximum energy so far detected. Furthermore, photons have to escape from regions that might be opaque to gamma-rays because of pair production, unless special conditions are satisfied. The peak gamma-ray isotropic luminosity turns out to be $L_{\gamma} \approx 3 \cdot 10^{36}$ erg s$^{-1}$ above 100 MeV. It is unknown whether this emission is leptonic or hadronic. The site and ultimate origin of this extreme particle acceleration depend on the disk-corona dynamics which lead to jet formation and relativistic propagation. A number of complex phenomena such as coronal magnetic field reconnection or shock acceleration along the proto-jets can take place and influence the formation of particle energy distribution functions up to GeV kinetic energies or beyond. In a leptonic scenario, synchrotron radiation in the GeV domain requires large magnetic fields near $10^4$ G and Lorentz factors $\gamma \sim 10^5$ or larger. Alternately, inverse Compton emission at GeV energies on the soft X-rays from the disk and/or optical/IR radiation from the companion star also requires Lorentz factors $\gamma \sim 10^5$ before being suppressed by the Klein-Nishina regime at TeV energies. Depending on the geometry (the Cyg X-3 jet inclination angle is constrained[7] to be near 10º-20º), gamma-ray emission can be boosted or partially suppressed. Alternately, hadronic scenarios involve proton-proton interactions (in the shock-accelerated front interacting, e.g., with the Wolf-Rayet mass outflow) in relatively dense environments. In this case, protons have to be accelerated to Lorentz factors $\gamma \sim 10^2$ or larger, and the interaction may require a critical target density to occur.

Complex as it may be, the theoretical modelling of Galactic microquasars such as Cyg X-3 requires particle acceleration and non-thermal photon production up to GeV energies as crucial ingredients. Until now, only one fast episode of TeV emission has been detected[26] from Cygnus X-1. The occurrence of extreme particle acceleration at special Cyg X-3 spectral transitions suggests that even higher energy emission up to TeV energies may be detected by future fast-response microquasar observations.

## TABLE CAPTION

Table 1 – Major gamma-ray flares of Cygnus X-3.

Column (1) shows the dates of the gamma-ray flares from Cyg X-3. In all cases, Cygnus X-3 showed a low-intensity or non detectable hard X-ray flux above 20 keV.

Column (2) gives the X-ray state as determined by the public data in the band (2-12 keV) of the All Sky Monitor (ASM) on board of RXTE.

Column (3) gives the radio flux state at the time of the gamma-ray flare.

Column (4) shows the time delay $dT_1$ between the gamma-ray flare and the radio state minimum, if applicable.

Column (5) provides the magnitude of the major radio flare following the gamma-ray flare.

Column (6) gives the time delay $dT_2$ between the gamma-ray flare and the major radio flare.

Column (7) gives the gamma-ray flare photon flux above 100 MeV. The errors include the statistical and take into account the effects of the Galactic diffuse emission and the presence of nearby sources.

## FIGURE CAPTIONS

**Fig. 1** – Hard X-ray daily flux of Cyg X-3 (in counts/sec in the energy range 15-50 keV) as a function of time (given in Modified Julian Days, MJD) as monitored by the BAT instrument onboard of the *Swift* satellite from January 1, 2008 until June 25, 2009 (http://heasarc.gsfc.nasa.gov/docs/swift/results/transients).

The red arrows mark the dates of major gamma-ray flares of Cyg X-3 as detected by the AGILE instrument above 100 MeV and reported in Table 1 (April 16-17, 2008; Nov. 3-4, 2008; Dec. 20-21, 2008; Jun. 22, 2009). The orange arrow also marks a low-intensity gamma-ray flare (Jun. 20-21, 2008). Grey colored areas show the time intervals of good AGILE gamma-ray exposure of Cyg X-3 with off-axis angles less than 45 degrees.

**Fig. 2** – Multifrequency data of Cyg X-3 during the period 13-27 April 2008. (*Upper panel*:) the major radio flare of Cyg X-3 detected by the RATAN-600 radio telescope SAO RAS. Peak radio emission above 10 Jy at 11 GHz was detected[20] on April 18, 2008 (MJD 54574). (*Second panel*:) simultaneous AGILE gamma-ray monitoring of the Cyg X-3 region. A major gamma-ray flare from a source consistent with Cyg X-3 above 100 MeV is detected $dT_2 \sim 1$ day before the measured radio peak time. The gamma-ray upper limits are at the 2-sigma level. (*Third panel:*) X-ray flux of Cyg X-3 as monitored by the ASM of XTE in the energy band 1.3-12.1 keV. The hardness ratio F(5-12 keV)/F(3-5 keV) increases by a factor of 3 on Apr. 17, 2008. (*Bottom panel*:) hard X-ray flux of Cyg X-3 as monitored quasi-continuously by Super-AGILE in the energy range 20-60 keV for an average of 14 observations/day.

Fig. 3 – Schematic representation of the radio and X-ray spectral states of Cygnus X-3 (adapted from ref. 14). The red stars mark the approximate positions of the major gamma-ray flares detected by AGILE (see Table 1) that *all* occur during pre-quenched and/or pre-flare radio states. These “gully” states are rare in Cyg X-3, and occur only in a few percent of the total time[8,14] (see Suppl. Information). For all gamma-ray flaring episodes, the radio and hard X-ray flux levels are low or very low, whereas the X-ray flux (1-10 keV) is large.

**Table 1 – Major gamma-ray flares of Cygnus X-3**

| Gamma-ray flaring date | X-ray state | radio state | d $T_1$ (days) | following radio flare | d $T_2$ (days) γ-ray/radio | γ-ray flux $10^{-8}$ ph. $cm^{-2}$ $s^{-1}$ (E > 100 MeV) |
|---|---|---|---|---|---|---|
| (1) | (2) | (3) | (4) | (5) | (6) | (7) |
| 16-17 Apr. 2008 (MJD = 54572-54573) | soft | pre-flare | | ~ 16 Jy (11 GHz) | ~ 0-1 | 260 +/- 80 |
| 2-3 Nov. 2008 (MJD = 54772-54773) | soft | pre-quenched | 3-4 | ~ 1 Jy (15 GHz) | ~ 8-9 | 258 +/- 83 |
| 11-12 Dec. 2008 (MJD = 54811-54812) | soft | opt. thick-thin change | | ~ 3 Jy (11 GHz) | ~ 9-10 | 210 +/- 73 |
| 20-21 Jun. 2009 (MJD = 55002-55003) | soft | pre-quenched | ~4-5 | | | 212 +/- 75 |

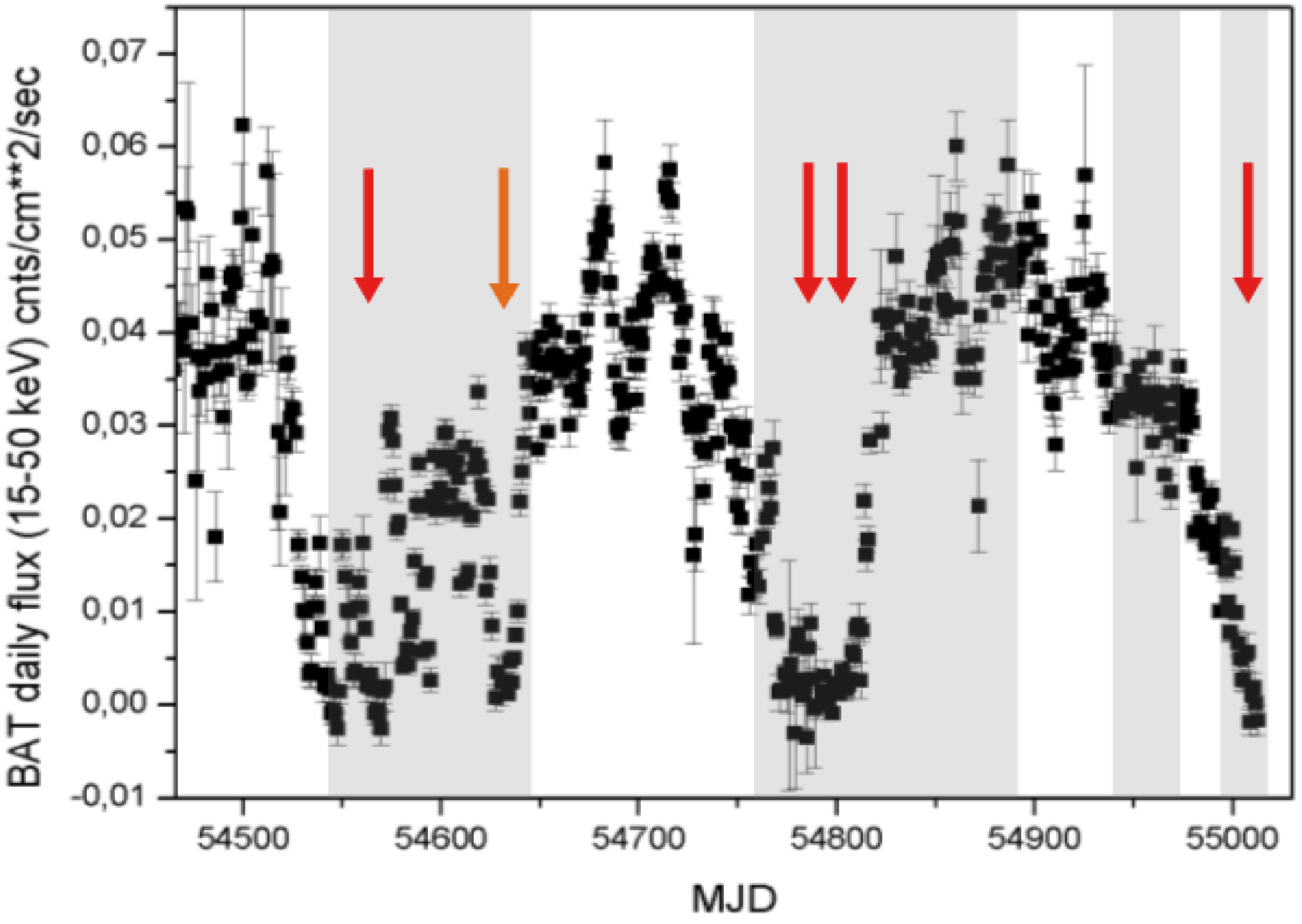

**Fig. 1 – Hard X-ray daily flux of Cygnus X-3.**

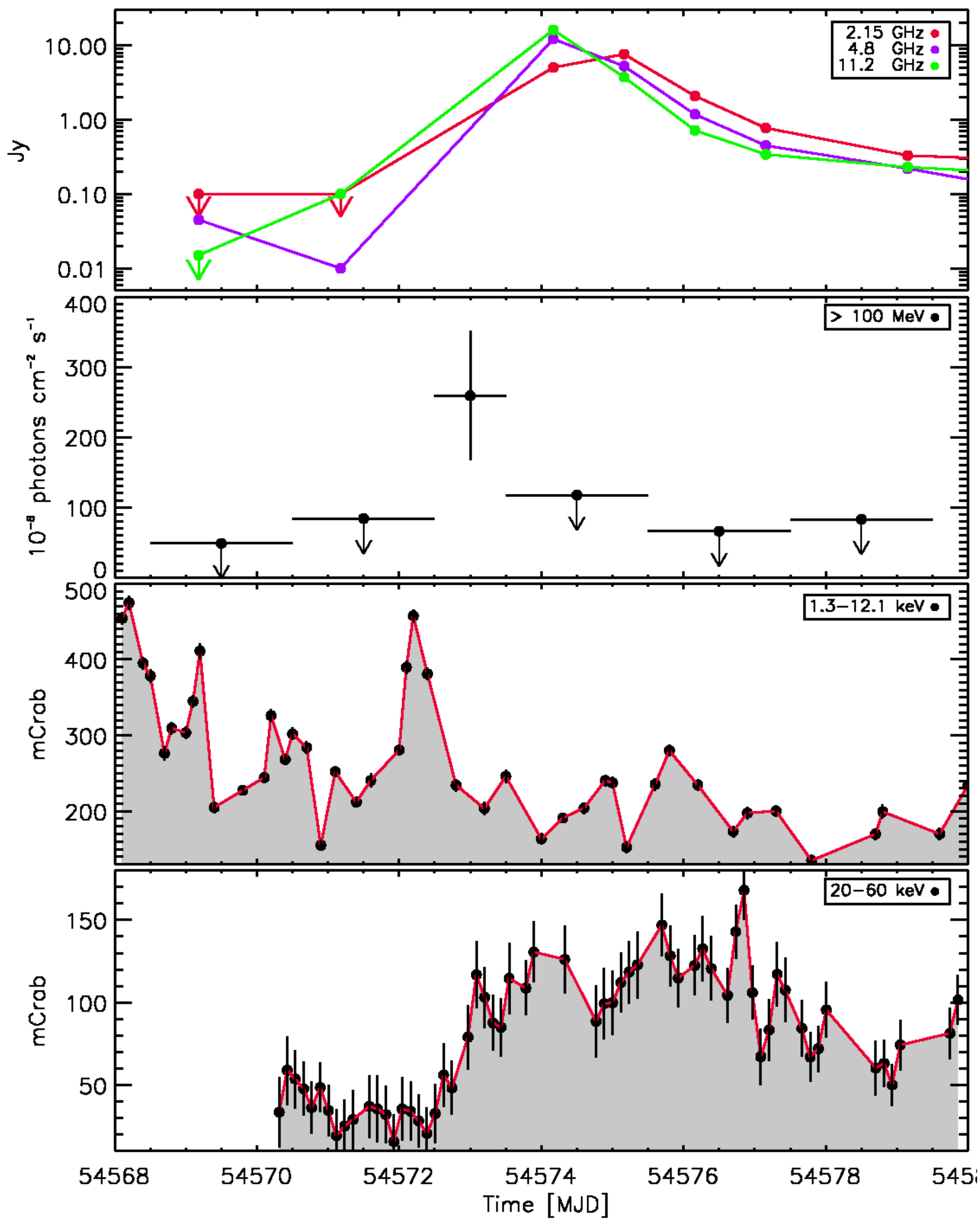

**Fig. 2 – Multifrequency data of Cygnus X-3 during the period 13-27 April 2008.**

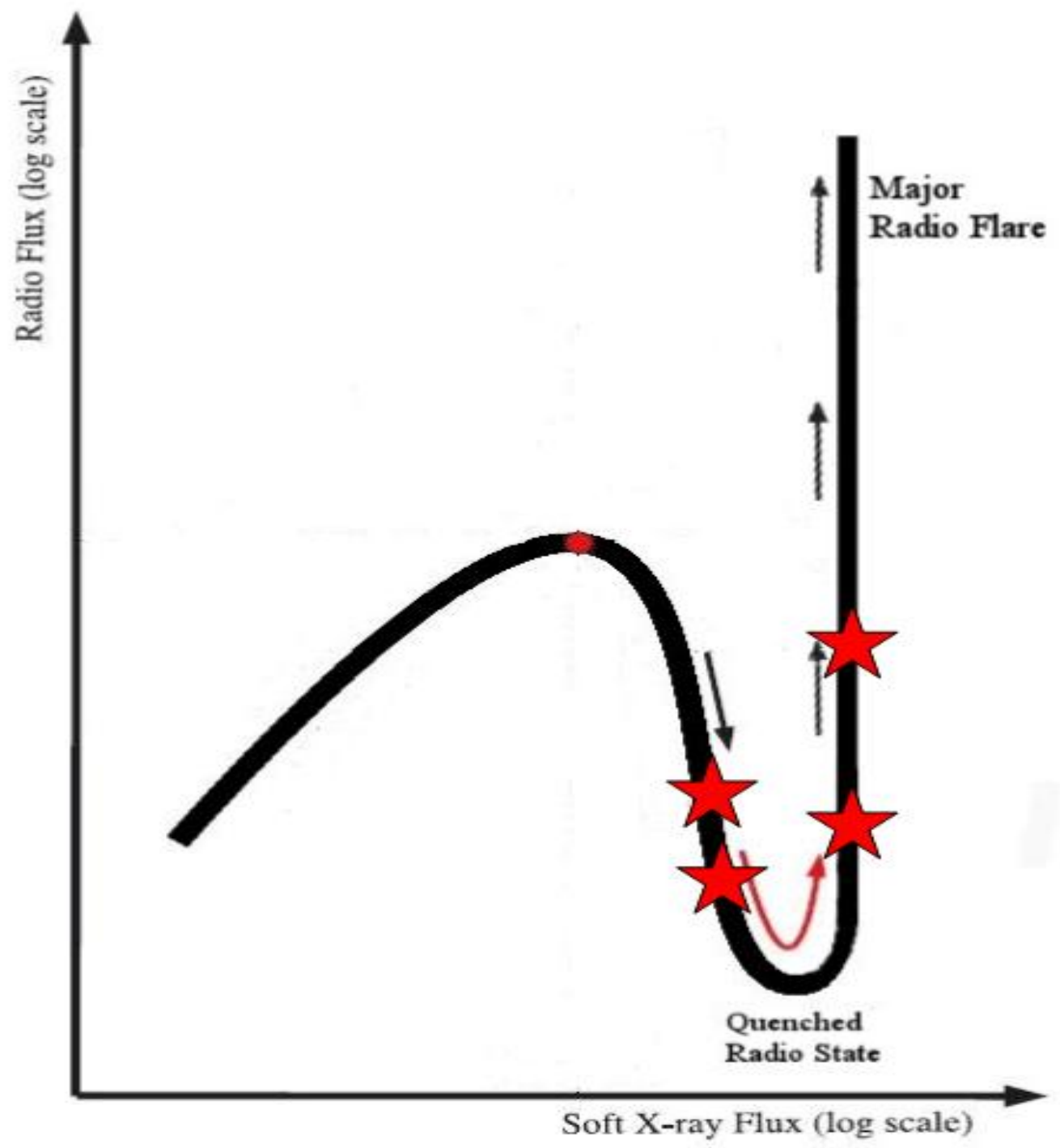

**Fig. 3 – Schematic representation of the radio and X-ray spectral states of Cygnus X-3.**

# Supplementary Information

In this section we provide additional information about the analysis of AGILE gamma-ray data of the Cyg X-3 region and of its gamma-ray flaring activity. AGILE[10,11] was launched in an equatorial orbit on April 23, 2007, and started scientific operations in July, 2007. Good sensitivity (for an on-axis effective area of about 400 $cm^2$ at 100 MeV), optimal angular resolution, and a very large field of view (FOV) of 2.5 sr are obtained at gamma-ray energies for a low particle background. The AGILE gamma-ray imager (GRID) can reach source location accuracy of 0.1-0.2 degrees in the 100 MeV-10 GeV energy range for deep integrations, and 0.5-1 degree location accuracy for 1-2 day timescale integrations. The hard X-ray imager (Super-AGILE) can monitor hard X-ray sources within a 1 sr field of view with arcminute angular resolution and a typical flux sensivity of about 10-15 mCrab. We carried out our analysis of gamma-ray data with the standard AGILE-GRID FM3.119_2 and FT3ab_2 filters and with the MULTI-2 likelihood analysis package available at the AGILE Data Center and Cycle-2 Guest Observers. We used the AGILE standard model[16] of Galactic diffuse gamma-ray emission.

Fig. 4 shows the gamma-ray intensity map above 100 MeV of the Cygnus Region obtained by integrating all available data obtained for off-axis angles less than 45 degrees. Several gamma-ray sources are clearly detected as confirmed by our multi-source maximum likelihood analysis (MSLA). Three most prominent gamma-ray sources (included in the AGILE[27] and *Fermi*-LAT[17] Catalogs) need to be included in our MLSA of the Cyg X-3 field. They all turn out to be gamma-ray pulsars, and their Galactic coordinates (in degrees) and average fluxes above 100 MeV are:
1AGL J2022+4032 (F=120 x $10^{-8}$ ph. $cm^{-2}$ $s^{-1}$, l = 78.29, b = 2.09, LAT-PSR J2021+4026, ref. 17),
1AGL J2021+3651 (F= 70 x $10^{-8}$ ph. $cm^{-2}$ $s^{-1}$, l = 75.28, b = 0.15, AGILE-PSR J2021+3651, ref. 28),
1AGL J2032+4102 (F= 40 x $10^{-8}$ ph. $cm^{-2}$ $s^{-1}$, l = 80.14, b = 0.97, LAT-PSR J2032+4102, ref. 17).

We are interested in the region near Cyg X-3, at Galactic coordinates (l = 79.85, b = 0.70). Our refined analysis of the integrated intensity, that supersedes that of ref. 27, shows two sources within a radius of 1° from the Cyg X-3 position.

The brightest of the two is a persistent nearby gamma-ray source worth of attention for its closeness to Cyg X-3. This source is detected both by AGILE (1AGL J2032+4102, with refined position: l= 80.14, b= 0.97) and by *Fermi*-LAT (0FGL J2032.2+4122; l = 80.2, b = 1.0), with a gamma-ray flux above 100 MeV of F = (40 +/- 6) x $10^{-8}$ photons $cm^{-2}$ $s^{-1}$. Pulsed gamma-ray emission from 0FGL J2032.2+4122 has been recently detected[17] by *Fermi*-LAT with a period P = 0.143 sec. We include this steady gamma-ray source in the analysis of this paper, and insert it permanently in the multisource maximum likelihood analysis with a constant flux and centroid position given by (l = 80.2, b = 1.0).

A second source is positionally consistent with Cyg X-3 and shows up by integrating all available data. We find a weak (4.6 sigma) gamma-ray source consistent with the Cyg X-3 position at the position (l, b) = (79.91, 0.46) + 0.37 (stat.) + 0.10 (syst.) and flux F = (10 +/- 2) · $10^{-8}$ ph. $cm^{-2}$ $s^{-1}$ above 100 MeV.

We searched over the whole dataset for *transient* gamma-ray emission in the Cyg X-3 region using the maximum likelihood analysis and False Discovery Rate[18,19] method. An automatic search was implemented using the AGILE-GRID multi-source likelihood analysis that can simultaneously fit several sources taking into account the diffuse Galactic background model. We searched over the whole database restricting the analysis only to good quality data and off-axis angles less than 45 degrees. A total effective time of approximately 5 months (corresponding to a gamma-ray exposure above 100 MeV near $10^9$ $cm^2$ sec) was accumulated in the Cyg X-3 region with off-axis angles less than 45°. We selected only events above 3 sigma as determined by the GRID-MSLA by including all persistent gamma-ray sources in the Cygnus region as specified above. The results are given in Table 2.

For each flaring event (shown in Fig. 5), we determined the centroid position, 90% confidence level error box, flux above 100 MeV and 400 MeV, the confidence level of the detection as determined by the MSLA and by the False Discovery Rate (FDR) statistical method[18,19] based on count statistics and local background fluctuation analysis. For all events reported in Tables 1 and 2 we determined the consistency with the Cyg X-3 position (given the 1-2 day integrations and photon statistical and systematic errors) and the overall statistical significance of the detection. We note that the FDR-$\alpha$ parameter provides the *fraction of expected false detections* for a given source selection. This method is complementary and more conservative than the determination of a "post-trial" detection significance based on a simulated set of replicated photon maps. Applying our FDR method to typical 1-day integrated AGILE-GRID photon maps in the Galactic plane, FDR-$\alpha$ values near or below 0.01 fully qualify for gamma-ray transients. We performed extensive numerical simulations of gamma-ray counts maps and studied the Poisson random fluctuations comparing them with the AGILE-GRID data. We find that values of FDR-$\alpha$ values near or below 0.01 correspond to post-trial random occurrences of 1-day map replications equal to 1 every 300 or more. Values of FDR-$\alpha$ ~ 0.001 or below turn out to be even rarer.

After more than a decade of radio and X-ray monitoring of Cyg X-3 we can determine the time occurrence of the special radio/X-ray states which are coincident with the gamma-ray flares. From Ryle/AMI Large Array radio data at 15 GHz, the source spends 2.9% of its time at fluxes below $F_{15GHz}$=0.01 Jy, and 1.6% below $F_{15GHz}$=0.005 Jy. From GBI data taken at 2.25 and 8.3 GHz during the period MJD 50409-51823 the time occurrences of radio fluxes with $F_{2.2GHz}$ < 0.015 Jy and $F_{8.3GHz}$ < 0.015 Jy are 1.4% and 1.2%, respectively. From the *Swift*-BAT data, for the period MJD 53413-55043, the time occurrences of hard X-ray states with $F_{15\text{-}50\ keV}$ < 0.001 cts $cm^{-2}$ $s^{-1}$ or undetectable flux are 6.8% and 3.3%, respectively. Thus we obtain a first estimate of the average time probability $< p_1 > = 0.03$ for Cyg X-3 to be in a radio/hard X-ray quenched state. Additional and crucial information is provided by the simultaneous soft X-ray flux. From the RXTE/ASM single dwell-data for the period MJD 50087- 55035 we obtain an occurrence of 2.7 % and 0.9 % for X-ray fluxes $F_{1.3\text{-}12.1keV}$ > 35 cts/s (467 mCrab) and $F_{1.3\text{-}12.1keV}$ > 40 cts/s (533 mCrab), respectively. We can consider the probabilities of the *combination* of simultaneous hard X-ray and soft X-ray fluxes at the times of the gamma-ray flares. For three out of four flares of Table 1 (April 16-17, 2008; November 2-3, 2008; June 21-22, 2009), the occurrence probabilities for the combination of the hard/soft X-ray states are in the range $0.003 < p_2 < 0.016$. The fourth gamma-ray flare of Dec. 11-12, 2008 has $p_2 = 0.03$. We obtain an average value of $< p_2 > = 0.016$.
Thus we determine two different estimates of the average probability of occurence of the set of four gamma-ray flares detected by AGILE in coincidence with the special Cyg X-3 states, $P_1 = (< p_1 >)^4\ N \approx 2.4 \cdot 10^{-4}$ and $P_2 = (< p_2 >)^4\ N \approx 2 \cdot 10^{-5}$, where N=300 is the total number of exposure days (trials).

Considering that all these flare detections were obtained for standard off-axis angles well below 45 degrees, we obtained the cumulative counts, exposure, and intensity maps by summing all four major flaring episodes of Table 1. By integrating all flaring data, we find that a gamma-ray source is detected at 5.5 sigma level at the average Galactic coordinate location (l,b): ($79^o.6$, $0^o.5$) +/- $0.5^o$ (stat.) +/- $0.1^o$ (syst.), with an average flaring flux F = (190 +/- 40) x $10^{-8}$ photons $cm^{-2}$ $s^{-1}$ above 100 MeV. After barycentric correction and orbital period folding, no statistically significant periodic gamma-ray signal was found.

We briefly comment here on the major gamma-ray flaring episodes of Cyg X-3.

**April 16-17, 2008**

This gamma-ray event occurred about 1 day before the strongest radio flare of Cyg X-3 during the period 2008/mid-2009. Fig. 2 shows the multifrequency lightcurve of Cyg X-3 at radio, gamma-ray, X-ray and hard X-ray energies during the week centered at the major radio flare on April 18, 2008. Several important facts can be deduced by our data: (1) the timing of the gamma-ray flare coincides with a 1-day peak of the soft X-ray emission during a period of low-flux emission at hard X-ray energies; (2) no gamma-ray emission is detectable in coincidence with the radio flare peak and subsequent slow decay. As clearly showed by Super-AGILE data of Fig. 2, Cyg X-3 changed state immediately following the soft-X-ray/gamma-ray peak producing a prominent hard X-ray increase together with a major radio flare probably followed by relativistic plasmoid ejections (similar to what recorded in ref. 7). The system changed from a soft X-ray to an intermediate hard X-ray state in a few hours, as shown by the detailed Super-AGILE lightcurve.

**November 3, 2008**

A 1-day gamma-ray flare was detected[29] by the AGILE-GRID on Nov. 2-3, 2008, with Cyg X-3 in a prolonged soft X-ray state with apparently no peculiar high-energy characteristic. This flare appears to be relatively "soft" in gamma-rays compared to the other flaring episodes. A more careful investigation of the radio lightcurve as monitored at the AMI Large Array[25] (AMI-LA) shows that the gamma-ray flare occurred when Cyg X-3 was just entering a quite deep "quenched radio state" that shortly thereafter reached a minimum of 1.5 mJy at 15 GHz (ref. 22). Such a quenched radio state of Cyg X-3 at this level of radio emission is relatively rare and was not detected during the previous months of monitoring at AMI-LA. This state was followed a few days later (as common[8,14] for Cyg X-3) by a major radio flare reaching about 1 Jy at 15 GHz. Fig. 6 (upper panel) shows the radio lightcurve at 15 GHz of early November 2008. Once again, the gamma-ray flare near 100 MeV occurred at a peculiar transition in preparation of a major radio flare. We note that *Fermi*-LAT, for different source livetime, photon selection, and daily exposure near 100 MeV compared with AGILE, does not detect significant emission during the same period[30]. We confirm this result by integrating the emission above 100 MeV with the relatively hard band-pass filter FM3.119-2.

**December 12, 2008**

The prolonged soft state (with undetectable hard X-ray emission by BAT/Super-AGILE) lasting about a month changed in mid December, 2008 into a hard state. This transition coincided with a major radio flare that occurred on December 20, 2008 after a very interesting evolution of the radio flux as shown in Fig. 6. The AGILE-GRID monitored the Cygnus region throughout the whole November-December, 2008 period. Weak gamma-ray emission is detected since the beginning of December, 2008. From Fig. 6 we notice that the radio flux increases around Dec. 1, 2008. A most significant gamma-ray flare in the Cyg X-3 region occurred on December 12, 2008. An analysis of the RATAN-600 data at different frequencies shows that this gamma-ray flare coincided with a transition of the radio emission from optically thick to optically thin states, that could provide evidence for the final expansion of a radio blob at the beginning of the strong radio flare reached a few days later. Fig. 7 shows the details of the multifrequency monitoring of Cyg X-3 for the period centered on the Dec. 12, 2008. The gamma-ray flare occurred during a prolonged low-level hard X-ray emission and in coincidence with a soft X-ray peak (in a way resembling the April 16-17, 2008 episode). A few days later, the hard X-ray flux substantially increased reaching a large level during the surge of the radio flux near Dec. 20, 2008.

**June 21-22, 2009**

The last major gamma-ray flare from Cyg X-3 that occurred on June 21-22, 2009 is unusual with respect to the others for two reasons: (1) the hard X-ray emission was not quiescent or on the verge of rising but rather decreasing, as clearly shown in Fig. 1; (2) the event was apparently not followed a few days later by a major radio flare, even though the sparse radio monitoring at the AMI Large Array cannot exclude the presence of a fast major radio flare that might have escaped detection. Fig. 8 shows the 15 GHz radio lightcurve of Cyg X-3 and the timing of the gamma-ray flare that might have occurred at the beginning of a quenched radio state, as it is suggested by the data trend.

## TABLE CAPTION

Table 2 – Gamma-Ray flares of Cygnus X-3: a detailed analysis.

Column (1) shows the dates of the major gamma-ray flares from Cyg X-3.
Column (2) gives the positioning in degrees (Galactic coordinates) obtained by the AGILE-GRID multi-source likelihood analysis (MSLA). The used filter is indicated in Column 4.
Column (3) gives the gamma-ray photon flux above 100 MeV.
Column (4) gives the MSLA significance of the detection. Note that the presence of the nearby gamma-ray source (1AGL J2032+4102/LAT-PSR J2032+4102) and the diffuse Galactic emission of the region makes the confidence levels obtained by the MSLA relatively low. Used GRID filters (FM for FM3.119_2 and FT for FT3ab_2) are specified in parenthesis. Column 5 provides an independent method of detection based on counts maps and proper background estimations.
Column (5) gives the False Discovery Rate[18,19] (FDR) $\alpha$ parameter, indicating the expected fraction of false detections for a give selection (for counts maps of the time intervals of Column 1). Values of FDR-$\alpha$ near 0.01 or below are typical of gamma-ray transients detected by the AGILE-GRID. They correspond to post-trial random occurrences of 1 every 300 1-day replications or more.

## FIGURE CAPTIONS

Fig. 4 - AGILE-GRID intensity map (in Galactic coordinates) above 100 MeV of the Cygnus Region ($8^o$x $5^o$.5 region) integrating all available data collected between November 2007 and June 23, 2009. The color bar scale is in units of $10^{-5}$ photons $cm^{-2}$ $s^{-1}$ $pixel^{-1}$. Pixel size is 0.1 degrees for a 3-bin Gaussian smoothing. The circles mark the positions of the prominent gamma-ray sources used in the AGILE-GRID likelihood analysis. The blue circles mark the 90% confidence contour levels of the AGILE positioning above 100 MeV of the sources 1AGL J2022+4032 (F=120 x $10^{-8}$ ph. $cm^{-2}$ $s^{-1}$, l = 78.29, b = 2.09, LAT-PSR J2021+4026, ref. 17), and 1AGL J2021+3651 (F=70 x $10^{-8}$ ph. $cm^{-2}$ $s^{-1}$, l = 75.28, b = 0.15, AGILE PSR J 2021+3651, ref. 28). The black circles mark the corresponding *Fermi*-LAT positionings. The Cygnus X-3 region appears to be quite complex in gamma-rays. We mark with a cyan circle the 90% confidence contour level of the AGILE positioning above 400 MeV of the source 1AGL J2032+4102 (F=40 x $10^{-8}$ ph. $cm^{-2}$ $s^{-1}$, l= 80.14, b= 0.97) that corresponds to the *Fermi* source LAT-PSR J2032+4102 (ref. 17). Cygnus X-3 (at the position l = 79.85, b = 0.70) is weakly detected by our MSLA above 100 MeV as a persistent source.

Fig. 5 – Most significant gamma-ray flaring activity detected by the AGILE-GRID above 100 MeV in the region near Cyg X-3 (marked with a black cross) for typical 1-day integrations (see Table 1). The panels show gamma-ray intensity maps above 100 MeV in Galactic coordinates of the field centered at Cyg X-3 and of area $8^o$x $6^o$.5 The color bar scale is in units of $10^{-4}$ photons $cm^{-2}$ $s^{-1}$ $pixel^{-1}$. Pixel size is 0.3 degrees for a 3-bin Gaussian smoothing. All maps were obtained with the FM3.119_2 filter except for the relatively “soft” event of Nov. 2-3, 2008 for which we used the FT3ab_2 filter. The green circles mark the circular approximations of the 90% confidence level error boxes taking into account systematic plus statistical errors for 1-day integrations. For the precise positioning and errors of the specific observations, see Table 2. The error boxes in cyan color mark the refined position of the nearby source 1AGL J2032+4102 as determined by AGILE above 400 MeV (outer contours) that is coincident with the *Fermi*-LAT source[17] 0FGL J2032.2+4122 (inner contours). This stable source is included in the multi-source likelihood analysis of the gamma-ray emission in the field.

Fig. 6 – Radio monitoring of Cygnus X-3 during the period November-December, 2008. (*Upper panel*:) radio lightcurve of Cyg X-3 at 15 GHz as monitored by the AMI Large Array (typical errors: 5% of the flux). During the first week of November, 2008 Cyg X-3 showed a prominent "quenched" radio state reaching ~ 2 mJy at 15 GHz. The red arrow mark the gamma-ray flare detected by AGILE on Nov.2-3, 2008 (entering into the radio quenched state), and on December 11-12, 2008. (*Lower panel*:) radio lightcurve of Cygnus X-3 at different frequencies as monitored by the RATAN-600 radio telescope. The blue arrow marks the time of the gamma-ray flare detected on Dec. 12, 2008 during a clear radio spectral state change from optically thick to optically thin states.

**Fig. 7 –** Multifrequency data of Cygnus X-3 during a 2-week period centered on Dec. 12, 2008. (*Upper panel*:) AMI-LA radio flux monitoring at 15 GHz. (*Second panel*:) AGILE-GRID gamma-ray lightcurve showing gamma-ray emission near Dec. 12, 2008. The gamma-ray upper limits are at 2-sigma level. (*Third panel*:) X-ray lightcurve as monitored by the ASM on board of XTE (1.3-12.1 keV). The hardness ratio (5-12 keV)/(3-5 keV) has a distinct peak on Dec. 12, 2008. (*Lower panel*:) hard X-ray flux lightcurve as monitored by BAT on board of *Swift* (15-50 keV).

**Fig. 8 –** Radio lightcurve of Cygnus X-3 at 15 GHz as monitored by the AMI-LA during the period May – June, 2009. During the second half of June, 2009, Cyg X-3 probably reached another "quenched" radio state at 15 GHz lasting for several days. The red arrow marks the gamma-ray flare detected by the AGILE-GRID that started monitoring again the region on June 15, 2009.

**Table 2 – Gamma-Ray flares of Cygnus X-3: a detailed analysis**

| Gamma-ray flaring date | γ-ray MSLA positioning (l,b) (in degrees) | γ-ray flux $10^{-8}$ ph $cm^{-2}s^{-1}$ | MSLA-σ | FDR-α |
|---|---|---|---|---|
| (1) | (2) | (3) | (4) | (5) |
| 16-17 Apr. 2008 (MJD = 54572-54573) | (79.1,0.6) ± 0.6 (stat.) ± 0.1(syst.) | 260 +/- 80 | 4.2 (FM) | 0.001 |
| 2-3 Nov. 2008 (MJD = 54772-54773) | (79.3,0.7) ± 0.7 (stat.) ± 0.1(syst.) | 258 +/- 83 | 4.0 (FT) | 0.01 |
| 11-12 Dec. 2008 (MJD = 54811-54812) | (79.6,0.3) ± 0.6 (stat.) ± 0.1(syst.) | 210 +/- 73 | 3.8 (FM) | 0.01 |
| 20-21 Jun. 2009 (MJD = 55002-55003) | (80.1,1.3) ± 0.6 (stat.) ± 0.1(syst.) | 212 +/- 75 | 3.6 (FM) | 0.001 |

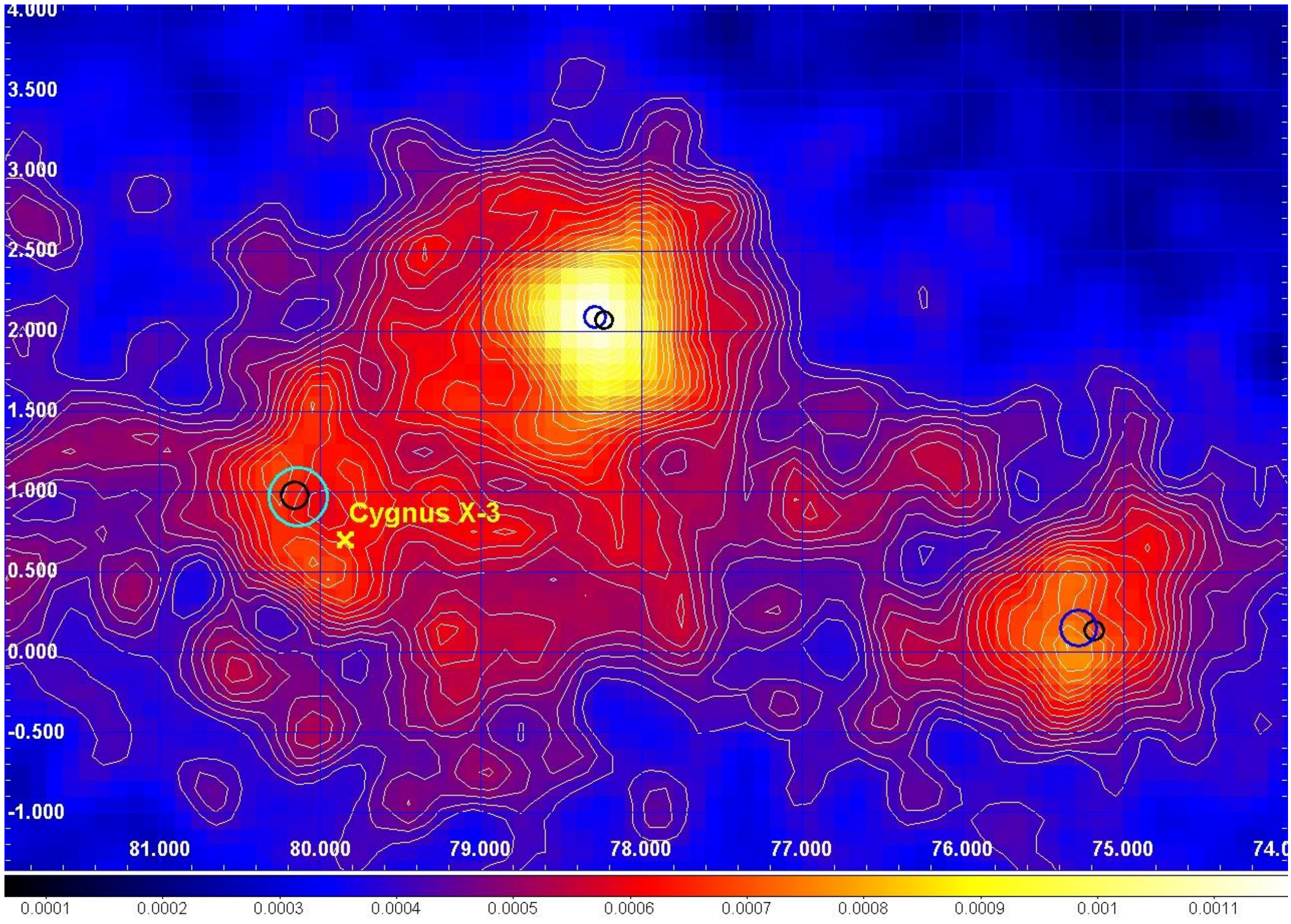

**Fig. 4 - AGILE-GRID intensity map (in Galactic coordinates) above 100 MeV of the Cygnus Region.**

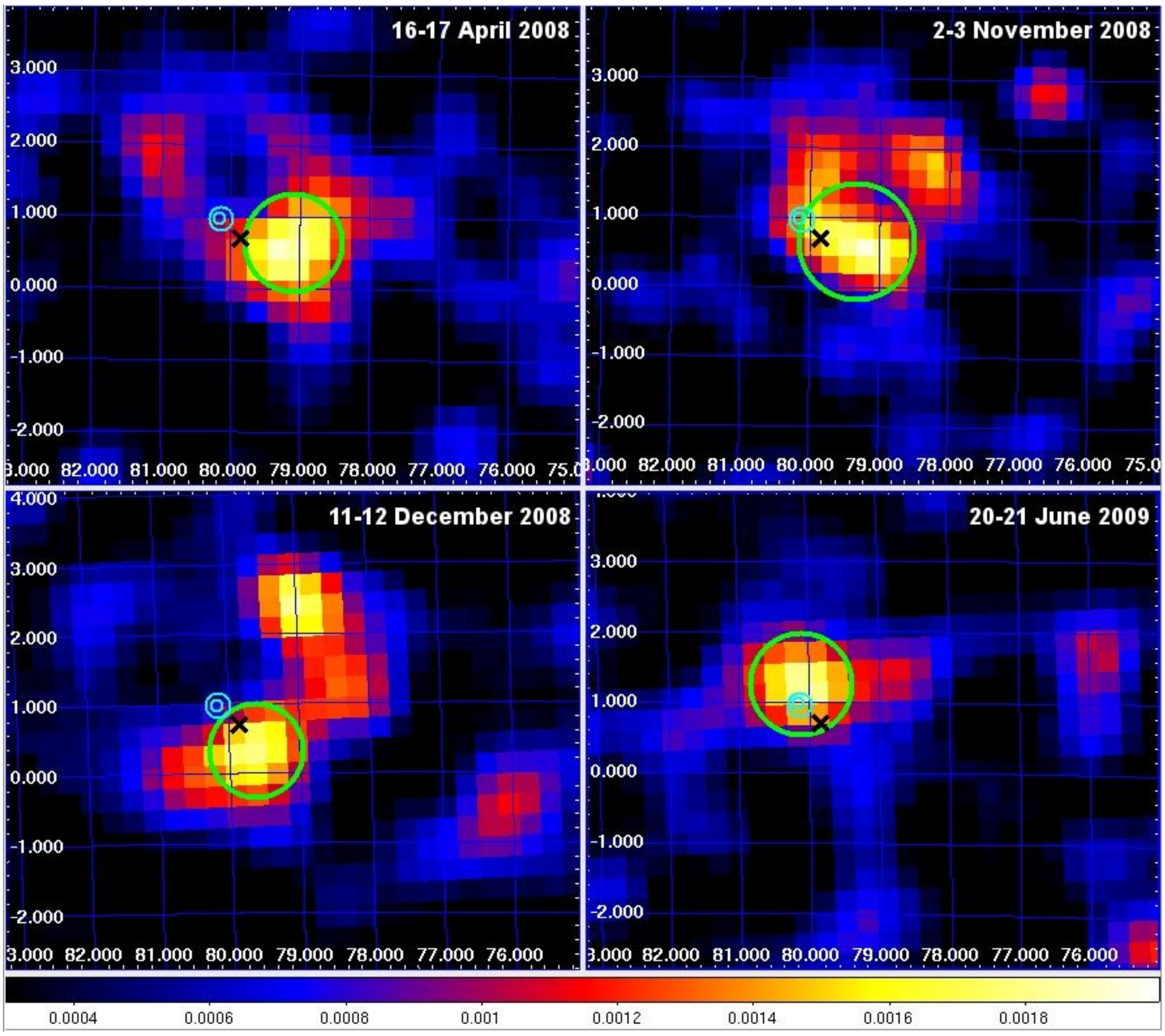

**Fig. 5 – Most significant gamma-ray flaring activity detected by the AGILE-GRID above 100 MeV in the region near Cygnus X-3.**

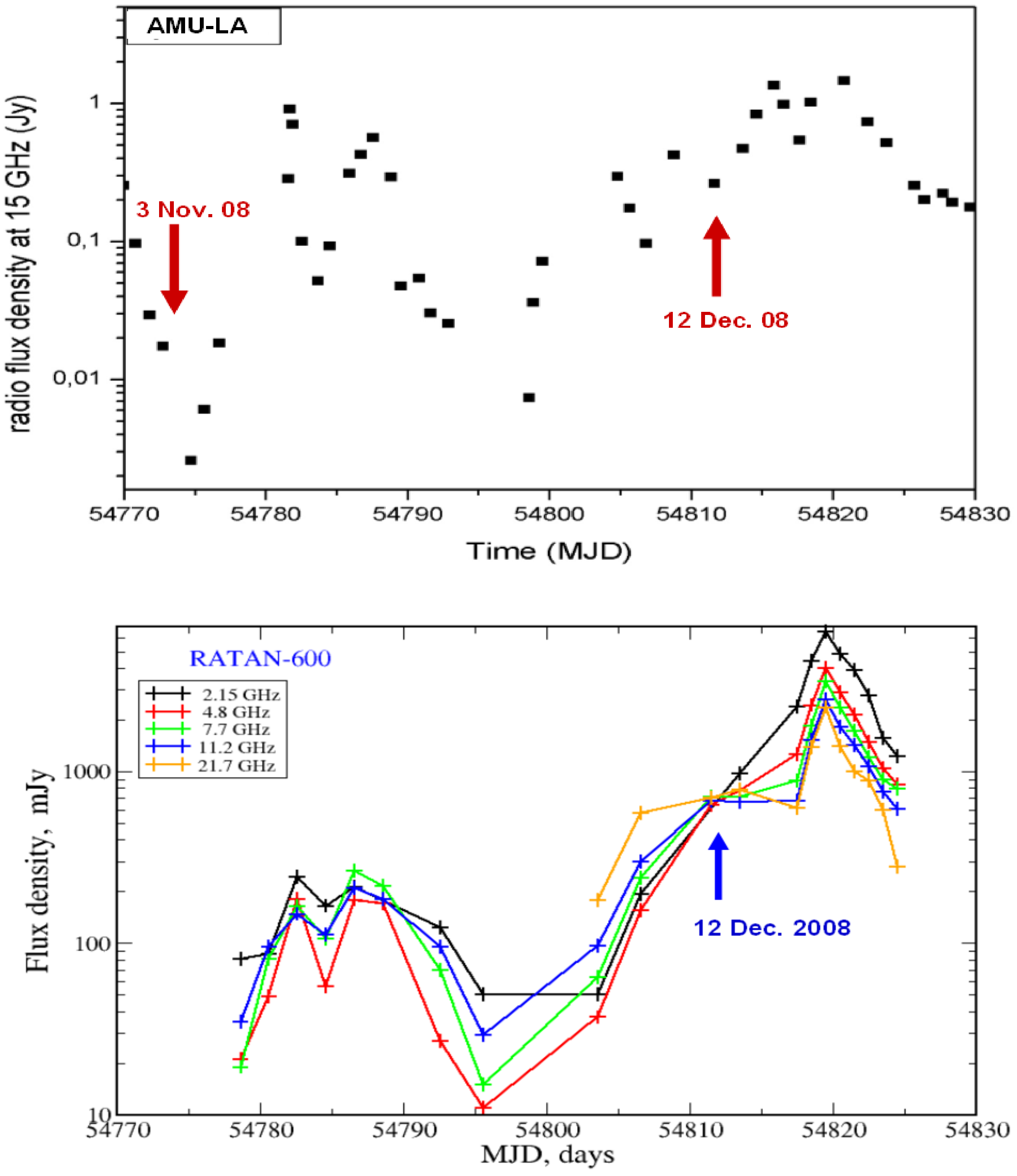

**Fig. 6 – Radio monitoring of Cygnus X-3 during the period November-December, 2008.**

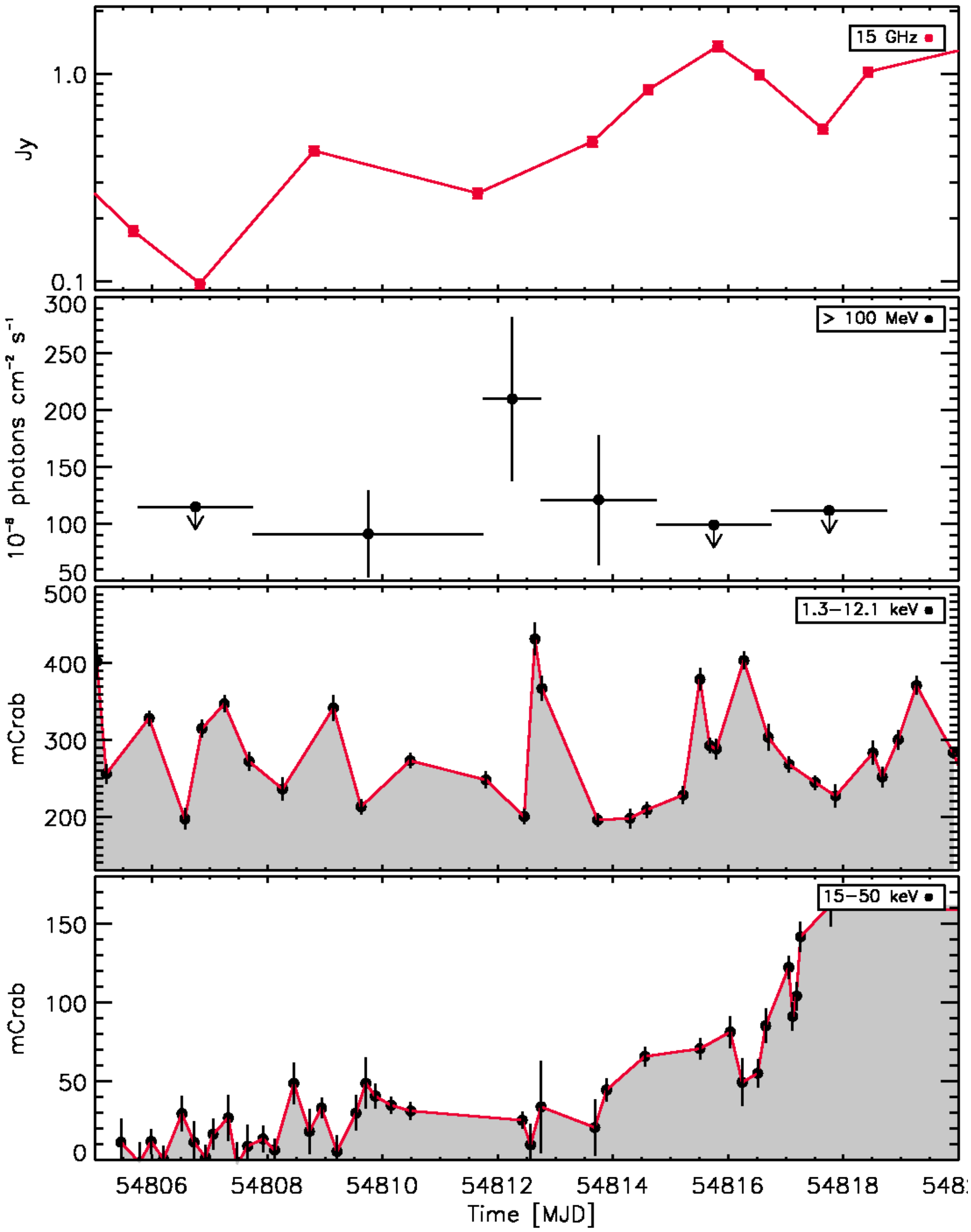

**Fig. 7 – Multifrequency data of Cygnus X-3 during a 2-week period centered on Dec. 12, 2008.**

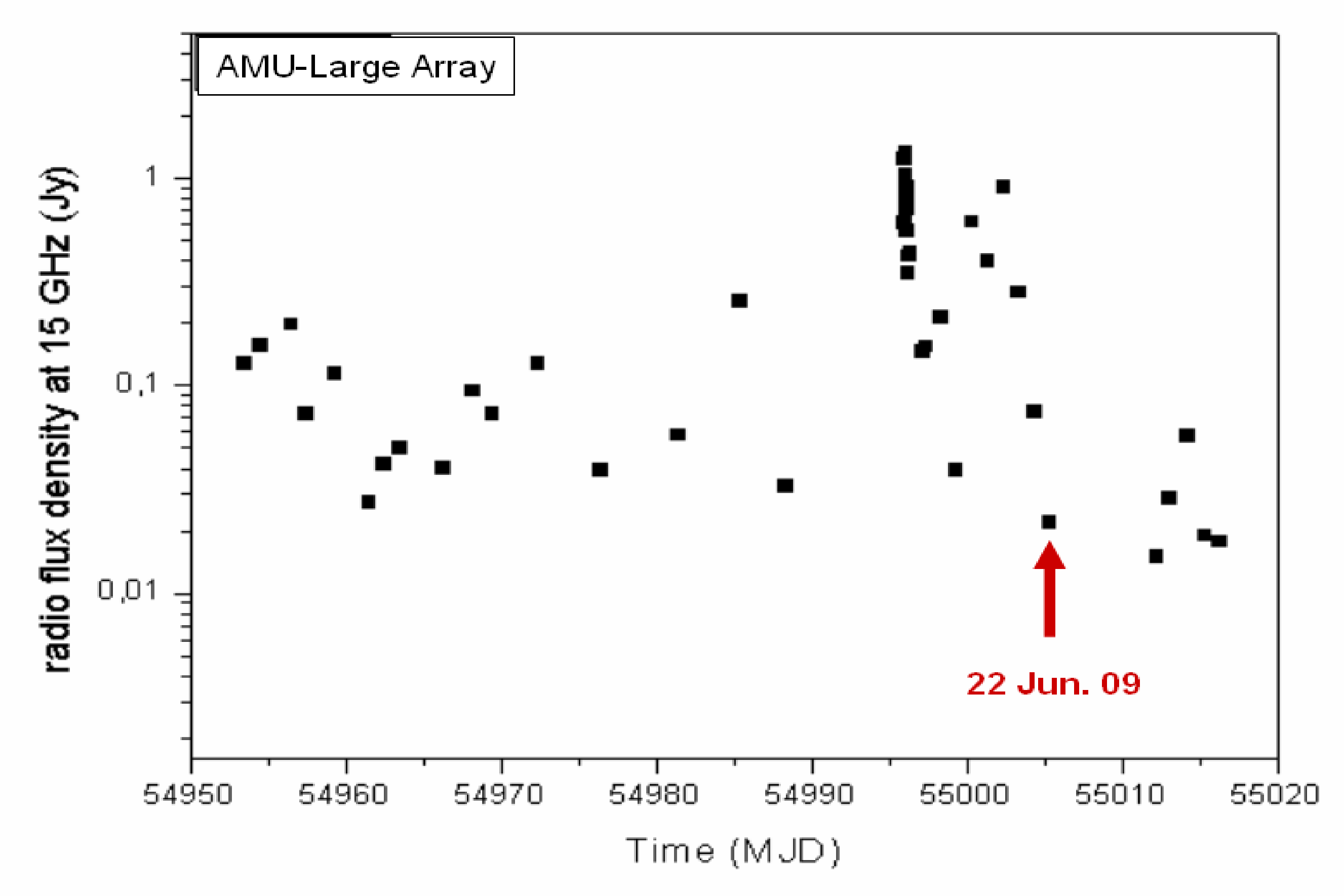

**Fig. 8 – Radio lightcurve of Cygnus X-3 at 15 GHz as monitored by the AMI-LA during the period May – June, 2009.**